\documentclass{aims}
\usepackage{amssymb,amsmath}
  \usepackage{paralist}
  \usepackage{graphics} %% add this and next lines if pictures should be in esp format
  \usepackage{epsfig} %For pictures: screened artwork should be set up with an 85 or 100 line screen
\usepackage{subfigure} %For pictures: screened artwork should be set up with an 85 or 100 line screen
 \usepackage[colorlinks=true]{hyperref}
\hypersetup{urlcolor=blue, citecolor=red}

\title[Recognition of crowd behavior]
      {Recognition of crowd behavior from mobile sensors with pattern analysis and graph clustering methods}

\author[Daniel Roggen, Martin Wirz, Gerhard Tr\"{o}ster and Dirk Helbing]{}


 \email{daniel.roggen@gmail.com}
 \email{martin.wirz@ife.ee.ethz.ch}
 \email{troester@ife.ee.ethz.ch}
 \email{dhelbing@ethz.ch}

\begin{document}
\maketitle

% Enter the first author's name and address:
\centerline{\scshape Daniel Roggen, Martin Wirz and Gerhard Tr\"{o}ster}
\medskip
{\footnotesize
% please put the address of the first author
 \centerline{Wearable computing laboratory, Gloriastrasse 35}
   \centerline{ETH Zurich}
   \centerline{CH-8092 Zurich, Switzerland}
} % Do not forget to end the {\footnotesize by the sign }

\medskip

\centerline{\scshape Dirk Helbing}
\medskip
{\footnotesize
 % please put the address of the second  and third author
 \centerline{CLU E11, Clausiusstrasse 50}
   \centerline{ETH Zurich}
   \centerline{CH-8092 Zurich, Switzerland}
}

%\bigskip
%
%% The name of the associate editor will be entered by an editorial staff
% \centerline{(Communicated by the associate editor name)}

%The abstract of your paper
\begin{abstract}
Mobile on-body sensing has distinct advantages for the analysis and understanding of crowd dynamics: sensing is not geographically restricted to a specific instrumented area, mobile phones offer on-body sensing and they are already deployed on a large scale, and the rich sets of sensors they contain allows one to characterize the behavior of users through pattern recognition techniques.

In this paper we present a methodological framework for the machine recognition of crowd behavior from on-body sensors, such as those in mobile phones.
The recognition of crowd behaviors opens the way to the acquisition of large-scale datasets for the analysis and understanding of crowd dynamics.
It has also practical safety applications by providing improved crowd situational awareness in cases of emergency.

The framework comprises: behavioral recognition with the user's mobile device, pairwise analyses of the activity relatedness of two users, and graph clustering in order to uncover globally, which users participate in a given crowd behavior.
We illustrate this framework for the identification of groups of persons walking, using empirically collected data.

We discuss the challenges and research avenues for theoretical and applied mathematics arising from the mobile sensing of crowd behaviors.

\end{abstract}

%%%%%%%%%%%%%%%%%%%%%%%%%%%%%%%%%%%%%%%%%%%%%%%%%%%%%%%%%%%%%%%%%%%%%%%%%%%%%%%
% INTRODUCTION   INTRODUCTION   INTRODUCTION   INTRODUCTION   INTRODUCTION
%%%%%%%%%%%%%%%%%%%%%%%%%%%%%%%%%%%%%%%%%%%%%%%%%%%%%%%%%%%%%%%%%%%%%%%%%%%%%%%
%\input{Introduction}

%%%%%%%%%%%%%%%%%%%%%%%%%%%%%%%%%%%%%%%%%%%%%%%%%%%%%%%%%%%%%%%%%%%%%%%%%%%%%%%
% INTRODUCTION   INTRODUCTION   INTRODUCTION   INTRODUCTION   INTRODUCTION
%%%%%%%%%%%%%%%%%%%%%%%%%%%%%%%%%%%%%%%%%%%%%%%%%%%%%%%%%%%%%%%%%%%%%%%%%%%%%%%
\section{Introduction}
\label{sec:introduction}

%There is nowadays a widespread distribution in the general population of interconnected on-body sensing and computing devices, in the form of mobile phones \cite{Campbell10}.
Nowadays, through the use of mobile phones, interconnected on-body sensing and computing devices are widespread in the population \cite{Campbell10}.
Processing signals from on-body sensor data can in principle reveal information about individual behaviors and activities \cite{Davies08} as well as collective human behaviors \cite{Pentland05}.
This ``reality mining'' \cite{Mitchell09} is deemed to provide objective measures of human interactions, also called ``honest signals'' \cite{Pentland08_book}.
This opens new ways in {\em Computational Social Science} \cite{Pentland_Science_2009}, where the vast amount of information obtained from mobile devices leads to new perspectives for sociometry and the analysis of social dynamics or crowd behavior.
To date, most work exploiting the information of body-worn sensors to sense collective aspects of human behavior have focused on social network analysis and the quantification of face-to-face interaction patterns.
Examples for this include the measurement of the statistics of friendship networks \cite{Pentland_pnas,Onnela07} and the improvement of social interactions in organizations \cite{Paradiso10,Buchanan07}.
%A few examples of this includes uncovering friendship networks from the communication flow identified from mobile phone data \cite{Pentland_pnas}, and the identification and improvement of social interaction in organizations \cite{Paradiso10,Buchanan07}.
The combination of data sources available online (e.g. through social network web sites) together with information from on-body sensors may, furthermore, lead to a convergence of social and technological networks \cite{Kleinberg08}.

\subsection{From modeling to sensing crowd behavior}

In our own work, we investigate the recognition of crowd behavior by analysis of data measured with on-body sensors.
The sensors are those typically used in today's mobile phones.
By crowd behavior, we understand the coordinated movement of a large number of individuals to which a semantically relevant meaning can be attributed, depending on the respective application.
Examples include a queue of people, the formation of uni-directional ``lanes'' in bi-directional pedestrian flows, the intersection of these lanes, or a group of people at a specific location.

The detection of such crowd behavior is a recent development.
It builds upon past work that extensively sought to understand the nature of collective and crowd behavior by means of models \cite{Helbing95,Coscia08,Bellomo08} and simulations \cite{Helbing00,Helbing05,Hoogendoorn03}.
For more details regarding modeling approaches, we refer to \cite{Zhan08}.

While the simulation of crowd behavior is a problem of synthesizing realistic pedestrian behaviors with a suitable model, the recognition of crowd behavior deals with the opposite problem.
It consists in selecting, among different competing models, the one that is most likely to explain the observed sensor data.
Thus, collecting sensor data from an ensemble of persons is a necessary condition to recognize crowd behaviors, but not a sufficient one.
The main challenge consists in the interpretation of this collected data and to devise methods to map the sensors signals, collected from an ensemble of persons, to one of several kinds of crowd behaviors.
This has to be done in a computationally efficient and robust manner.
To this end, statistical machine learning techniques play a key role \cite{Duda00}.
However, there is a wide range of other disciplines that are involved, such as information spreading, distributed data fusion, graph visualization, graph clustering, or network analysis.

The recognition of crowd behavior facilitates practical applications.
In situations of emergencies during large-scale public events, machine recognition of crowd behavior enables a better situational awareness of event managers and informed participants.
This may be used to carry out an evacuation more successfully and efficiently \cite{Helbing_encyclopedia}.
Measuring the dynamics of crowd behavior can also be useful for urban planning and pedestrian navigation.

The recognition of crowd behavior also promotes fundamental research.
Past insights into crowd behavior have mainly been obtained through video analysis (see e.g. Refs.~\cite{Saxena08,Johansson08,Zhan08,Johansson07}).
By relying on body-worn sensors, crowd behavior can be analyzed on a much larger scale, allowing experiments to take place under realistic everyday conditions.

\subsection{Contribution}

One of the main challenge in crowd behavior recognition is to infer the most likely crowd behavior underlying the data collected from an ensemble of persons.
We present in this article a methodological framework to address this problem. We refer to this framework as the {\em crowd behavior recognition chain}.
This framework does not define specific data processing methods.
It rather organizes a set of solutions to the problem of crowd behavior recognition along well defined processing stages.
This allows the structured comparison of various methods and allows to modularize the investigation of further crowd behavior recognition systems.
%In this paper we In particular, we propose here a combination of pattern analysis and graph clustering techniques for crowd behavior recognition.
% in the form of a crowd behavior recognition chain.

Since the sensing of crowd behavior is a recent development, we wish to explain based on this methodological framework, what are the applied mathematical fields that play a role.
On the one hand, crowd behavior sensing offers a new application domain for current methods.
On the other hand, that application domain has specific characteristics that will drive the development of new methods.
One particular challenge is to make sense of large amounts of data, which are likely to be ``noisy'' due to the high variability of human behaviors.

Let us now describe the principles that we have developed to recognize crowd behavior from body-worn sensors, emphasizing the families of applied mathematical methods that come into play at various stages of the data analysis.
Besides the specific methods used in our work, we mention other applied mathematical fields which offer relevant methods for the sensing of crowd behavior.
We eventually emphasize requirements to improve existing methods and to develop new ones.

Since our perspective is predominantly that of applying methods to empirically collected data, we take as an illustrative example the detection of human group formation from on-body sensors.
In order to emphasize the principles of our approach, we limit ourselves to rather simple mathematical methods. However, the reader will be able to make the link to more advanced methods within the same data analysis framework.

In particular, in this paper:
\begin{itemize}
	\item
	We briefly present the experimental and technical aspects underlying crowd behavior recognition (see sec. \ref{sec:experimental}), which consists in gathering on-body sensor data from an ensemble of persons.
	During system development, these persons follow an experimental protocol leading to pre-defined collective behaviors. Eventually, after the system has been optimized, it is used on the the data collected in everyday situations.
	\item
	We explain how the sensor data are analyzed (see sec. \ref{sec:arc}).
	This is first carried out for each person individually.
	The outcome is the identification of ``behavioral primitives'' and the quantification of their characteristics (hereafter refered to as ``individual behavior'').
	A pairwise measure of individual behavior relatedness is then computed, given the assumption that the two persons participate to a common crowd behavior.
	Finally, in order to obtain a global view of the structure of the crowd behavior from these pairwise measures, we use graph clustering to uncover which individuals participate in the same crowd behavior.
	Further network analysis techniques can be used to characterize the resulting graph.
	We formalize these methodological steps in the form of a ``crowd behavior recognition chain''.
	%The dominant approach consists of statistical machine learning and correlation analysis between behavioral primitives identified on pairs of subjects.
	%and Information spread???
	%\item The interpretation of the results may be envisioned with graph visualization and graph clustering techniques (section \ref{sec:graph}).
	The methods involved include essentially signal processing and statistical machine learning, similarity analyis, graph embedding and clustering, and network analysis.
	\item
	We illustrate the crowd behavior recognition for the example of walking as part of a group, based on experimental data acquired from 10 subjects performing various individual and collective activities (see sec. \ref{sec:application}).
	%The behavioral primitives and their characteristics identified on each person individually are then analyzed with a global perspective to uncover the common underlying process likely to explain the observed data (section \ref{sec:global}).
	\item We discuss the current challenges and the main fields of applied mathematics that are relevant in the analysis of large-scale data collected from on-body sensors to infer crowd behavior (see sec. \ref{sec:discussion}).
	\item Finally we conclude this paper emphasizing the benefits of on-body sensing for the analysis of crowd dynamics and its better understanding (see sec. \ref{sec:conclusion}).
\end{itemize}

%This approach is an extension of methods used by the ubiquitous computing community to recognize the behavior or activities of individual persons.
%
%
%
%
%mathematical methods of interest as well as the challenges to
%
%Our activities from the experimental and machine learning perspective.
%
%the challenges and
% approach we outlined for the recognition of crowd behaviors from on-body sensors.
%We
%

%\cite{Barabasi08}

%%%%%%%%%%%%%%%%%%%%%%%%%%%%%%%%%%%%%%%%%%%%%%%%%%%%%%%%%%%%%%%%%%%%%%%%%%%%%%%
% EXPERIMENTAL   EXPERIMENTAL   EXPERIMENTAL   EXPERIMENTAL   EXPERIMENTAL
%%%%%%%%%%%%%%%%%%%%%%%%%%%%%%%%%%%%%%%%%%%%%%%%%%%%%%%%%%%%%%%%%%%%%%%%%%%%%%%
%\input{experimental}

%%%%%%%%%%%%%%%%%%%%%%%%%%%%%%%%%%%%%%%%%%%%%%%%%%%%%%%%%%%%%%%%%%%%%%%%%%%%%%%
% INTRODUCTION   INTRODUCTION   INTRODUCTION   INTRODUCTION   INTRODUCTION
%%%%%%%%%%%%%%%%%%%%%%%%%%%%%%%%%%%%%%%%%%%%%%%%%%%%%%%%%%%%%%%%%%%%%%%%%%%%%%%
\section{Experimental aspects and datasets}
\label{sec:experimental}

\begin{figure}[htbp]
  \centering
  \subfigure[Illustration of the 3D acceleration sensor]{
    \label{fig:sensor}
			\includegraphics[width=.7\linewidth]{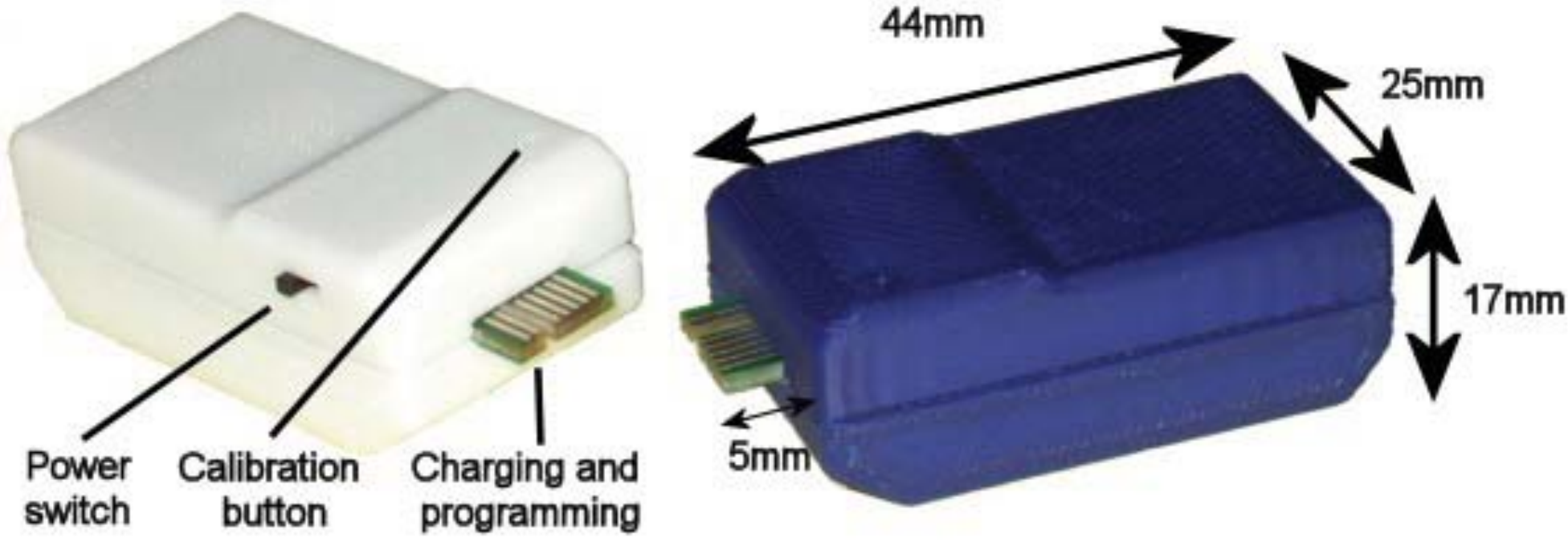}
	}
%  \subfigure[Untertitel 2]{
%    \label{fig:covovertime_energy}
%			\includegraphics[width=\linewidth]{./pics/covovertime_energy}
%	}	
	  \subfigure[Placing of the sensor on subject's hip]{
    \label{fig:sensorplacing}
			\includegraphics[width=.25\linewidth]{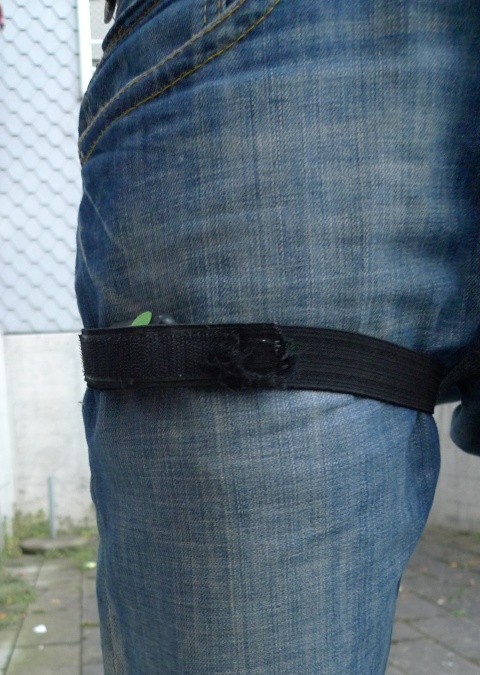}
	}	
  \caption{Acceleration sensor adjusted to the hip for the recognition of crowd behavior.
  The sensor placement mimicks that of a mobile phone located in the trouser pocket.
  The data acquired by the sensor are synchronized accross all participants of the experiment and stored for later analysis.}
  \label{fig:sensoroverview}
\end{figure}

\begin{figure}[!t]
\centering
\includegraphics[width=\linewidth]{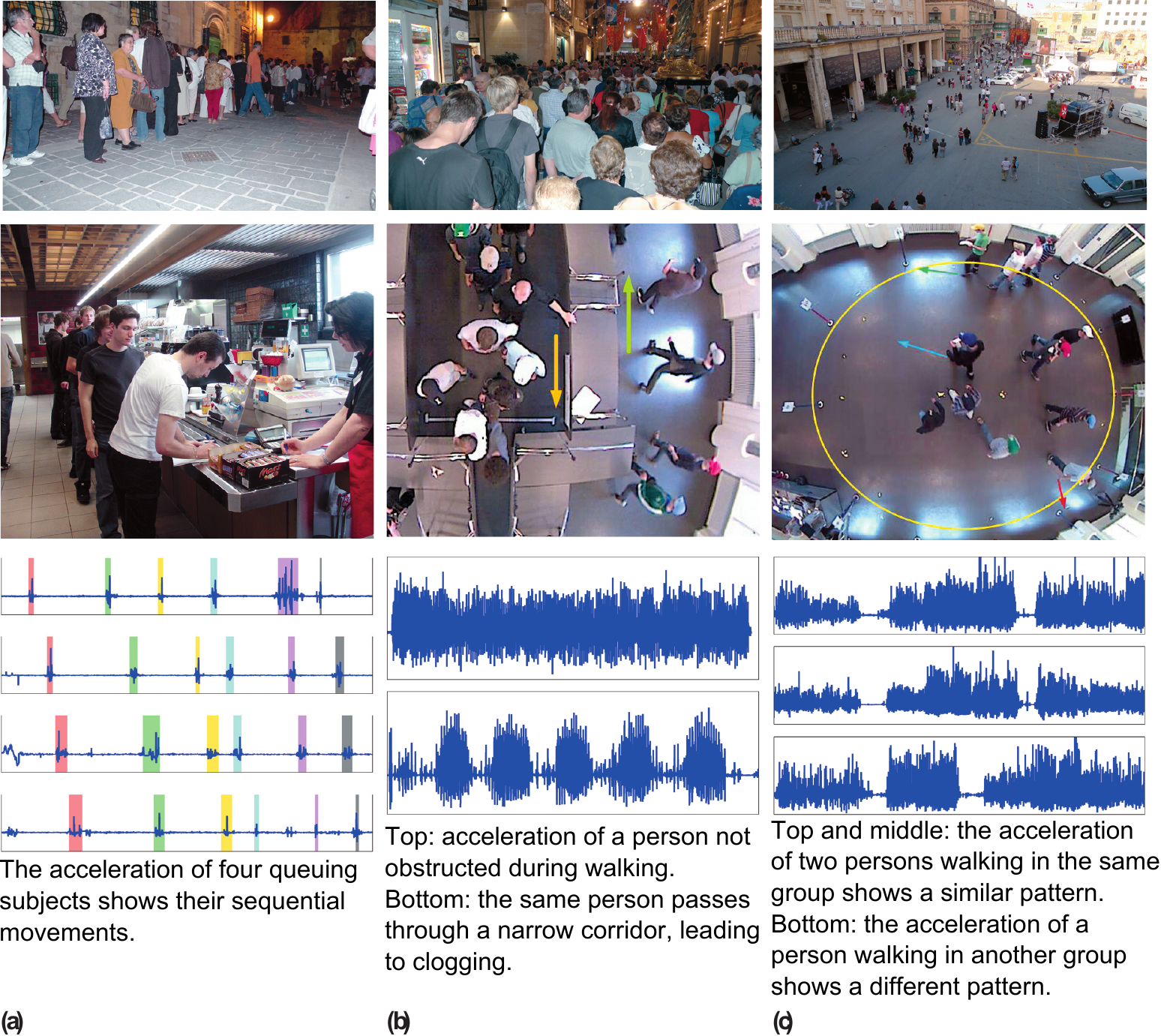}
\caption{Three typical crowd behavior: (a) queuing, (b) clogging, (c) group formation.
Starting with an identification of the collective behaviors of interests at large-scale events (top row), the relevant situation is emulated in the laboratory with a limited number of subjects wearing on-body sensors (middle).
Signals of accelerometers at the hips are depicted for different users (bottom).
The signals show specific patterns reflecting different collective behaviors, which we aim to automatically recognize in order to identify the current crowd behavior in which a person is involved.
}
\label{fig:experiments}
\end{figure}

The machine recognition of the activities of a single person has been extensively investigated in ubiquitous computing and wearable computing to create smart-assistants \cite{Mann98b}: systems that proactively and unobtrusively provide information or assistance upon detection of specific activities or gestures.
Examples of activity-aware systems include e.g.:
an industry worker's assistant that recognizes complex gestures of workers assembling a car \cite{Stiefmeier08}; gestural control of a computer \cite{Kallio06};
a system reacting to the freezing of gait of persons with Parkinson's disease  \cite{Baechlin10_TITB};
or systems to detect and prevent elderly from falling \cite{Benocci10} (see the special issue ``Activity-Based Computing'' for a survey of state-of-the-art approaches and applications \cite{Davies08}).
These systems operate by looking for patterns corresponding to specific activities in the data of various numbers of multi-modal sensors placed on the body of a single person.
This is usually tackled as a problem of {\em learning by demonstration} \cite{bao:04, Ward06,figo10preprocessing}.
In such a paradigm, the mapping between sensor signals and activities is learned using machine learning techniques \cite{Duda00}.
This is well suited for the complex and highly variable nature of human activities.
For this reason, we follow this approach for the recognition of crowd behavior.
A learning-by-demonstration approach is carried-out as follows for crowd behavior recognition (see ref. \cite{Wirz10a} for more details):

\begin{enumerate}
%	\item Identification of the collective behaviors from observational data (see figure \ref{fig:experiments}) or litterature (see e.g. \cite{Mataric93}).	 
	\item Participants are recruited to emulate in a laboratory setup the collective behavior of interest.
They are instrumented with sensors (see Figure \ref{fig:sensoroverview}).
They are instructed to perform a set of actions.
As a result of executing their tasks, a collective behavior occurs.
Thus, the collective behavior is generated under well-controlled conditions.
\item Data are collected synchronously from the sensors of all experimental subjects.
Supplementary data are obtained by ambient cameras.
These are used to identify the time points, at which the collective behavior of interest occurred (see Figure \ref{fig:experiments}).
\item The crowd behavior recognition chain is trained or optimized on a subset of the dataset ({\em training data}).
This includes optimizing the statistical models linking sensor signal characteristics to individual and collective behaviors, and optimizing all the other parameters of the system.
\item
The part of the dataset that was not used for training ({\em test data}) is used to estimate the performance of the crowd behavior recognition chain by cross-validation \cite{Duda00}.
\item
Finally, the system may be used on the data acquired in everyday situations.
At this point, ground truth is not available.
The crowd behavior recognition performance is assumed to be identical to what was obtained during the optimization process.
\end{enumerate}

Common sensors for human activity recognition are body-worn accelerometers and inertial measurement units (IMUs).
Accelerometers are extremely small and measure the vectorial acceleration of the device requiring very little power.
%, and many low-cost commercial solutions are available.
% for researchers (e.g. from SparkFun Electronics).
IMUs contain accelerometers, magnetometers and gyroscopes and allow one to sense the orientation of the device with respect to a frame of reference.
These sensors are nowadays found in mobile phones \cite{Campbell10}.
%Commercial solutions are provided by Xsens Technologies or InterSense.
%IMUs are typically placed on each body segment and allow to reconstruct a body model of the user.
%On-body microphones are also successfully used for activity recognition, as many human activities generate characteristic sounds (e.g. using a coffee machine, brushing teeth) \cite{Ward06}.Typical sensor modalities are listed in \cite{Roggen10_dataset}.
In Ref~\cite{Wirz10a} we review further sensors common in mobile phones that can be used for the recognition of crowd behavior.
In this article we present crowd behavior recognition results obtained from accelerometers.
These accelerometers can be provided by a standalone device, or by a smartphone.
Both kinds of devices deliver, for all practical purposes, identical acceleration patterns when placed at same body location (here on the front trousers pocket). Thus the results presented here are representative of those obtained with a broad set of mobile phones (e.g. Apple iPhone, Android Nexus One, and virtually any other smartphone containing an accelerometer, which is a broad portion of the current market).

In Figure~\ref{fig:experiments} we illustrate three kinds of collective behaviors, observed at a large-scale public event in the city of Malta.
These behaviors are people queuing, people clogging and forming lanes, and people walking in groups of small size.
In the rest of this article, we focus on the latter collective behavior: the identification of people walking together in a group.
Figure \ref{fig:experiments} illustrates the laboratory emulation of the observed behaviors with a limited number of subjects equipped with sensors.
Finally, the figure also shows the signals acquired from a sensor placed on the body. The sensor is an accelerometer worn on the hip as illustrated in Figure~\ref{fig:sensoroverview}.
One can visually observe characteristic signal patterns for each kind of collective behavior.
The objective for crowd behavior recognition is to identify these characteristic patterns automatically.
%Note that while the data presented here comes from a standalone acceleration sensor, a mobile phone

%%%%%%%%%%%%%%%%%%%%%%%%%%%%%%%%%%%%%%%%%%%%%%%%%%%%%%%%%%%%%%%%%%%%%%%%%%%%%%%
% ARC   ARC   ARC   ARC   ARC   ARC   ARC   ARC   ARC   ARC   ARC   ARC   ARC
%%%%%%%%%%%%%%%%%%%%%%%%%%%%%%%%%%%%%%%%%%%%%%%%%%%%%%%%%%%%%%%%%%%%%%%%%%%%%%%
%\input{arc}

%%%%%%%%%%%%%%%%%%%%%%%%%%%%%%%%%%%%%%%%%%%%%%%%%%%%%%%%%%%%%%%%%%%%%%%%%%%%%%%
% ARC   ARC   ARC   ARC   ARC   ARC   ARC   ARC   ARC   ARC   ARC   ARC   ARC
%%%%%%%%%%%%%%%%%%%%%%%%%%%%%%%%%%%%%%%%%%%%%%%%%%%%%%%%%%%%%%%%%%%%%%%%%%%%%%%
\section{Crowd behavior recognition chain}
\label{sec:arc}

\begin{figure}[htb]
  \centering
    \includegraphics[width=\linewidth]{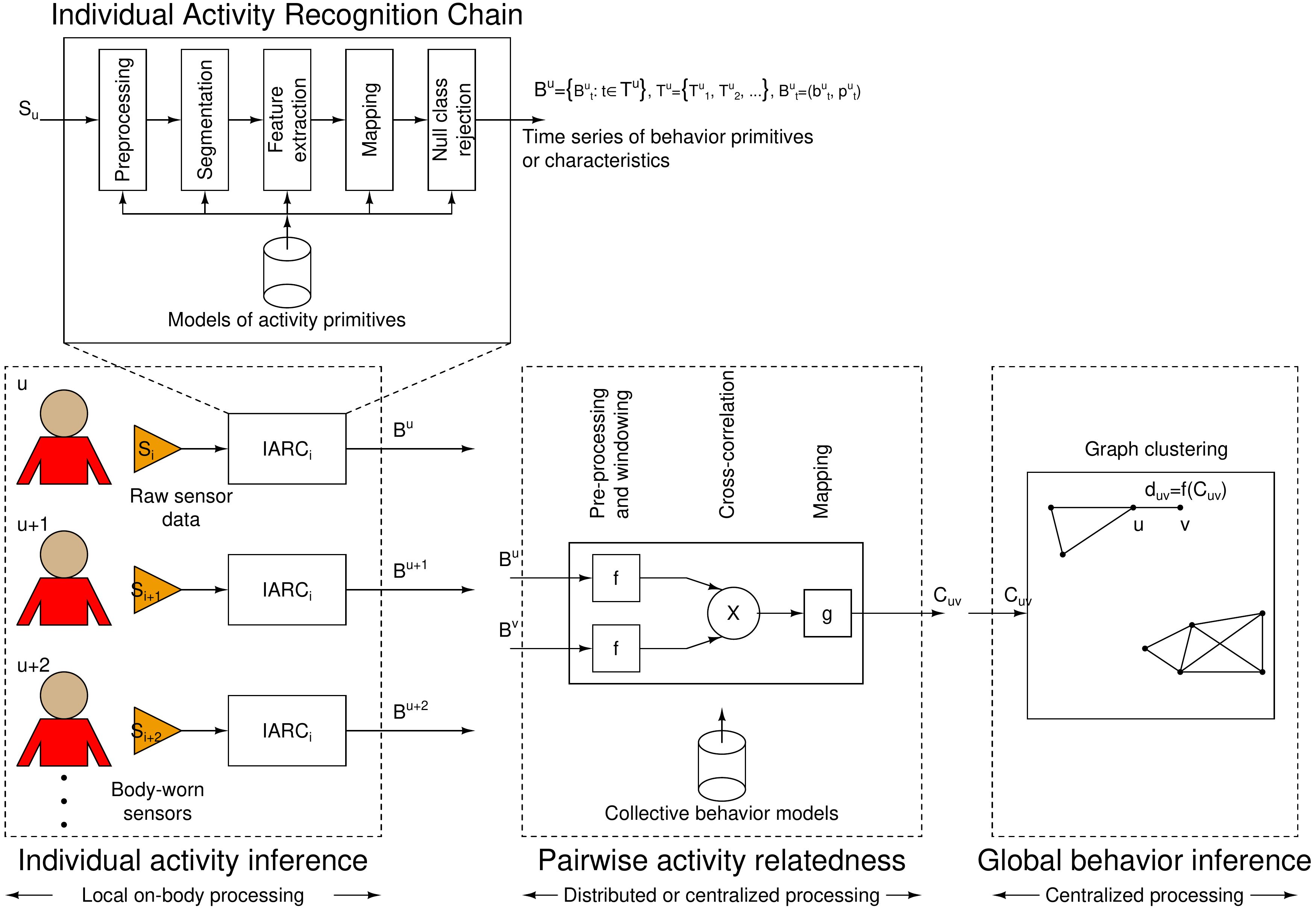}
	\caption{Processing steps used to infer from on-body sensor data, which users take part in a given crowd behavior.
	The data processing comprises three steps: First, characteristics of individual user behaviors are inferred, using the ``individual activity recognition chain''.
	This processing chain consists of signal processing and machine learning elements and maps sensor signals to time series of behavioral primitives or to quantitative characteristics of the user's behavior.
	Then, an analysis of the behavior of pairs of individuals is carried out, yielding a measure of disparity, in view of the assumed crowd behavior.
	Finally, graph clustering techniques are used to identify commonalities across all experimental subjects from the pairwise disparities.
	The resulting clusters identify the set of persons participating in a certain crowd behavior.
	}
	\label{fig:arc}
\end{figure}

We now formalize a series of processing steps (the ``crowd behavior recognition chain'') that can be used to infer crowd behaviors from body-worn sensors.
This defines a methodological framework.
The specific methods used within that framework may vary according to use cases.
The proposed process is illustrated in Figure~\ref{fig:arc}.
It consists of:
\begin{itemize}
	\item
	First, the crowd behavior recognition system is queried to identify the individuals that participate to a specific crowd behavior.
	This behavior is indicated by the user, according to the respective application.
	In other words, an assumption of the approach is that, among all individuals, some participate in a specified crowd behavior.
	The task of the recognition chain consists in identifying, which individuals participate to that crowd behavior.\footnote{This shares similarity, in a Bayesian framework, to the computation of the conditional probability of an observation (i.e. measured sensor data), given a phenomena (i.e. an assumed crowd behavior).}
	\item
	Then, characteristics of the behavior of each single person are inferred from the sensors worn by that person.
	Depending on the crowd behavior of interest, these characteristics may be ``activity primitives'' (e.g. performing a step) or a continous characterization of the user activity (e.g. the speed of walking).
	We refer to this as the ``individual behavior''.
	This is carried out with online signal processing and machine learning techniques.
	
	\item
	Then, the characteristics inferred previously are analysed pairwise for each pair of individuals to find out whether the behavior of these two persons may be the result from their participation to the specified crowd behavior.
	The outcome is a measure of disparity between two individuals as compared to the assumed crowd behavior.
	For example, two people of the same group are likely to walk, stop, and take turns at roughly the same time, and their locations are close to each other.

	\item Finally, the pairwise measures of disparity are analyzed by graph visualization and graph clustering.
	This allows one to determine which people, among all experimental subjects, take part in the common crowd behavior that the user of the system was interested in.
	Network analysis may then be used to characterize the structure of the crowd behavior in more details.
\end{itemize}

The above process is explained in more details in the following subsections.

The crowd behavior recognition chain contains a wide range of parameters.
These parameters are optimized during design time on the training dataset to correctly identify ensembles of people taking part in a common crowd behavior (see Section~\ref{sec:experimental}).
%Thus, there are two steps to the recognition of the crowd behavior.
%At design time, the parameters of the recognition chain are optimized with a dataset of crowd behaviors reserved for training.
%This training dataset contains ground truth annotations derived from cameras.
%Afterwards, the recognition chain is deployed in practical applications with the best identified parameters.
%At this point, ground truth is not made available.
%The performance of the recognition chain is assumed to be identical to what was obtained during the optimization process.
%This is ensured by using large scale training datasets and using cross-validation techniques \cite{Duda00}

\subsection{Individual activity recognition chain}
\label{sec:iarc}

\begin {table}[!hb]
	\centering
	{\scriptsize
		\begin{tabular}{|p{0.3\linewidth}|p{0.6\linewidth}|}
			\hline
			Individual behavior & The IARC detects or estimates \\
			\hline
			Mode of locomotion & The IARC detects the user's mode of locomotion and stance, such as walking, running, sitting, standing or lying. Individuals in a group are likely to have the same mode of locomotion. \\
			Stop and go & The IARC detects, when the user stops or starts walking. Individuals in a group are likely start and stop walking in a correlated manner.\\
			Foot stride & The IARC detects the foot steps of the user, such as the heel strike or toe-off. The phase of people walking as a group is likely to be synchronized.\\
			Turn &	The IARC detects sharp user turns (change in heading). People walking as a group are likely to show correlated turns.\\
			\hline
			\hline			
			Walking speed & The IARC indicates the walking speed of the user, which is more or less identical for people walking in a group.\\
			Heading & The IARC indicates the heading of the user, which is correlated for people walking in a group.\\
			\hline
		\end{tabular}
	}
	\vspace{0.5em}
	\caption{The individual activity recognition chain (IARC) is designed to recognize one or more user behaviors from the on-body sensor data, which are then processed to infer whether two persons participate in the same crowd behavior.
	This table indicates some user behaviors that are relevant for the recognition of walking in a group.
	The top part of the table describes sporadic behaviors (action primitives), which can be detected by the IARC.
	The bottom part indicates continuous behavioral characteristics which can be estimated by the IARC.
	These elements can all be recognized, using sensors available in today's mobile phones \cite{Shin10}.
	In this work, the IARC is designed to estimate the walking speed by computing, as a proxy, the variance of the user acceleration.
	Other crowd behaviors may require the identification of other kinds of behaviors.}
	\label{tbl:behaviors}
\end{table}

We refer to the {\em individual activity recognition chain} (IARC) as a set of processing principles, which is commonly followed by most researchers to infer human activities from raw sensor data \cite{Ward06,bao:04,figo10preprocessing,Bettini10} (see figure~\ref{fig:arc}, top).

The IARC maps low-level sensor data $S^u$ (e.g. body-limb acceleration) of user $u$ to semantically meaningful activity primitives (e.g. do a step, stand) or to quantitative characteristics of the activity (e.g. walking speed). We refer hereafter to this generically as the user ``individual behavior'' $B^u$.
Formally:
\begin{displaymath}
IARC: S^u \rightarrow B^u
\end{displaymath}

In Table~\ref{tbl:behaviors}, we indicate a few characteristics of the user behavior that are relevant for the detection of walking as a group.
Since the user behavior changes over time according to what he is doing, the system must identify individual behaviors as they occur sporadically (for activity primitives) or continuously (for a continuous characterization of the user activity).
This is realized by streaming signal processing and statistical machine learning techniques in real time.
This online data processing consists at a given time of estimating the individual behavior, using the data available up to that time point.\footnote{The amount of data taken into consideration prior to the time at which the system takes a decision usually depends on the duration of the behaviors of interest.
For instance, if one is interested in detecting steps, the amout of data considered would correspond roughly to the duration of a step (for instance one second, for normal walking).}

Therefore the behavior $B^u$ corresponds to a time serie:
\begin{displaymath}
B^u = \{B^u_t : t \in T^u \},
\end{displaymath}
where $T^u=\{T^u_1, T^u_2, \ldots \}$ represents the time instants $T^u_t$ at which the system takes a decision about the behavior $B^u_t$ of the user $u$.
The behavior can be represented by a tuple $B^u_t = (b^u_t,p^u_t)$, where $b^u_t$ is either a set of discrete activity primitives (e.g. performing a step, stopping, turning), or a continous characterization of the activity (e.g. the walking speed). $p^u_t$ optionally represents the confidence of the system in the decision, which may be exploited in a Bayesian framework.

``Meaning'' is attributed to the sensor data streams by ``comparing'' them with known parameterized activity prototypes through a series of processing stages that are internal to the IARC.
The IARC processing stages are (see Fig.~\ref{fig:arc}):\footnote{We omit the superscript $u$ for legibility reasons.}
\begin{itemize}
	\item
{\em Sensor data acquisition}.
	A time series corresponding to the sensor data is obtained: $S = \{\boldsymbol{s}_0, \boldsymbol{s}_1, \boldsymbol{s}_2, ...\}$.
		Since sensors can provide multiple values (e.g. an acceleration sensor provides a 3D vector), and multiple sensors can be jointly sampled, a vectorial notation is used.

\item
{\em Signal pre-processing}.
	The time series $S$ leads to a pre-processed time series $P$:
	\begin{displaymath}
	P = \{\boldsymbol{p}_0, \boldsymbol{p}_1, \boldsymbol{p}_2, ...\}
	\end{displaymath}
	Typical transformations are calibration, de-noising, or a computation of the acceleration amplitude from the vectorial acceleration.

\item
{\em Segmentation} of the time serie $\boldsymbol{P}$ into sections of interest (i.e. a subset of $\boldsymbol{P}$), which are those likely to contain an activity primitive (e.g. for footstep detection) or a time span within which a characteristic of the user behavior is computed (e.g. for the walking speed).
	The section $i$ is delimited by a start time $t^s_i$ and an end time $t^e_i$ within the time series $\boldsymbol{P}$, yielding a segmented time series $W_i$:
	\begin{displaymath}
	W_i = \{\boldsymbol{p}_{t^s_i}, ..., \boldsymbol{p}_{t^e_i}\}
	\end{displaymath}
	A common type of segmentation technique is the sliding window, for periodic movements such as walking.
	In that case, the window has a fixed size $w_1$, and we have $t^s_i = t^e_i-w_1$.
	
\item
{\em Feature extraction.}
	Features are computed on the before mentionned sections to reduce their dimensionality and to discriminate activities of interest.
	The result is a feature vector $\boldsymbol{X}_i$:
	\begin{displaymath}
	\boldsymbol{X}_i = \Psi( W_i )
	\end{displaymath}

\item
Next, a {\em mapping} of the feature vector $\boldsymbol{X}_i$ into an individual behavior $b_i$ is performed:
	\begin{displaymath}
	\boldsymbol{X}_i \rightarrow b_i, p_i
	\end{displaymath}
	This mapping can be a machine learning classifier, to detect sporadic activity primitives, in which case $b_i$ is one of a discrete set of possible individual behaviors.
	The mapping can also be a transformation which yields a continuous-valued characteristic of the user behavior $b_i$.
	Usually, the classification yields the likelihood of the detected activity primitive $p_i$.
	%These are often represtented by probabilities $p_i$ with Bayesian approaches, and
	In fact, many classifiers can be calibrated to provide probabilistic outputs \cite{Cohen04}.
	%Other measures include the Dempster-Schafer degree of belief, or the degree of agreement between multiple classifiers or multiple sensors.
	
%{\em Decision fusion} combines multiple information sources (multiple sensors, or multiple classifiers operating on one sensor) into a decision about the activity that occurred.
	
\item
{\em ``Null-class'' rejection}.
	In cases where the confidence in the classification result is low, the system may discard the classified activity $i$ based on $p_i$.	

\end{itemize}

The outcome is a time series of tuples $\{(b^u_i, p^u_), \ldots\}$, reflecting that the individual behavior $b_i$ is detected with likelihood $p_i$ at times $t^e_i$.

%Note that when continuous characteristics of the behavior such as walking speed are desired, the classification into a discrete set of output classes is usually replaced by a mapping from the feature vector $\boldsymbol{X}_i$ to the continuous characteristic of interest (e.g. the walking speed is usually proportional to the energy of the acceleration signal).

The parameters of the IARC include the thresholds to segment activities or reject the null class, the set of features, scaling parameters, and the classifier parameters.
They are determined at design time using a training dataset $D$ that contains data instances (feature vectors) $\boldsymbol{X}$ and corresponding ground truth labels $\gamma$ indicating to which individual behavior they correspond:
\begin{displaymath}
D = \{ (\boldsymbol{X}_i,\gamma_i) \}_{i=1}^{N}
\end{displaymath}

%Before operation, the classifiers used in the ARC are trained using a training set containing data instances (feature vectors) $\boldsymbol{X}$ and the corresponding activity label $\gamma$.
%Other parameters, such as the thresholds to segment activities or reject the null class, or the set of features, are also optimized prior to operation.

Classifiers commonly used for the recognition of modes of locomotion, such as walking, running, sitting, standing, etc, have been reviewed in Ref.~\cite{figo10preprocessing}, together with typical features derived from acceleration signals.
The classifiers that are typically used include Support Vector Machines \cite{Qian10}, decision trees, k-Nearest Neighbors or Naive Bayes classifiers \cite{randell:00}.
%This is commonly used with periodic activities when frequency domain features are used (e.g. walking leads to energy in specific frequency bands).
If the temporal unfolding of the sensor signal must be analyzed,
%, such as the identification of the heel-strike and toe-off events during gait,
approaches such as Dynamic Time Warping \cite{Ko05} or hidden Markov models (HMMs) \cite{Starner98b} are used.

Note that the IARC does not guarantee a perfect recognition of the individual activities.
Data fusion approaches can further enhance the IARC accuracy by combining several on-body sensors, e.g. with ensemble classifiers \cite{Zappi08a}. Higher-level activity models, reasoning and probabilistic approaches can also be included to further enhance the IARC performance \cite{ranganathan_reasoning_2004,mckeever09}.
Nevertheless modes of locomotion can be recognized with accuracies above 80\% independently of the on-body sensor placement \cite{bao:04} and other simple activities are nowadays also well recognized.

\subsection{Local, pairwise disparity analysis}
\label{sec:arc_disparity}

A measure of disparity between a pair of individuals $u$ and $v$ is inferred at time $T$ from the behaviors $B^u$ and $B^v$.
The disparity value reflects the possibility that the individual behaviors of persons $u$ and $v$ are the consequence of their participation to a common crowd behavior that the system must recognize.

Note that the disparity could also be computed directly between a pair of sensor signals $S$ instead of the behaviors $B$.
However, such a direct correlation between individual sensor readings may not be sufficient to recognize the crowd behaviors.
This may be due to noise (e.g. resulting from different on-body sensor placement and orientation, which affect the readings of a 3D accelerometer), variable latency of the wireless data transmission, or the need to use multimodal sensing to capture the individual activities that are required to infer a desired crowd behavior. For instance speech features and body movement may be used to identify the positive or negative emotional state of the user. Together with the data of other persons, this may be used to identify whether a person is part of a cheerful crowd or a sad crowd, for instance at a stadium during a match. Direct sensor correlation may not work due to the highly diverse way in which persons can express their emotions and the fact that they are likely not expressing them synchronously.
Thus, the role of the IARC is to evidence individual activity patterns that are, in combination with those of other persons, highly discriminative of the crowd behavior to recognize.
Nevertheless, since the crowd behavior recognition chain is a methodological framework, it also allows for a direct correlation of the sensor data, if this is desired: in that case the IARC would simply do no processing and $B^u=S^u$.

The disparity is computed as follows:
\begin{displaymath}
C^{u,v}_T = g(Corr(f(B^u,T),f(B^v,T)))
\end{displaymath}

The functions $f$, $Corr$ and $g$ depend on the specific crowd behavior that is considered.
Typically, $f$ is a pre-processing function.
$Corr$ computes a measure of similarity between the input data, with respect to the crowd behavior of interest.
Finally, $g$ maps this to a disparity value.
We define the disparity to be 0 if the behaviors of two users are likely resulting from their participation to the same crowd behavior.
Conversely, the disparity tends to 1 if the behaviors are not likely to be the result of participating to the same crowd behavior.
This yields the disparity matrix $\boldmath{C_T} = [C^{u,v}_T]_{n \times n}$ at time $T$.

The disparity is usually computed in real-time with a sliding window technique.
Thus, $f$ defines a temporal window $w_2$ within which the disparity is computed, considering only the behavior $B^u_t$ for $t \in T^u, T-w_2 \leq t \leq T$:
\begin{displaymath}
f(B^u,T) = \{B^u_t : t \in T^u, T-w_2 \leq t \leq T \}
\end{displaymath}

%The similarity $Corr$ is often a correlation function and $g$ normalizes the output value range.
%The pre-processing $f$, the nature of the correlation $Corr$ and the mapping $g$ of the correlation to a disparity value are determined by an optimization process at design time based on the training dataset.
The functions $f$, $Corr$, and $g$ as well as their parameters are determined by an optimization process at design time based on the training dataset.
This process aims at obtaining lower values $C^{u,v}_T$ when two individuals $u$ and $v$ take part in the same crowd behavior at time $T$, and higher values $C^{u,w}_T$ when two individuals $u$ and $w$ do not take part in the same crowd behavior.

%Typically, at a given time $T$, we seek for two persons $u$ and $v$, who take part in the same crowd behavior to lower values of $C^{u,v}_T$ than the value $C^{u,w}_T$ for two persons $u$ and $w$ that do not take part in the same crowd behavior.
%The functions are selected and their parameters are optimized during design time based on the training datasets.

\subsection{Inferring the global crowd behavior}
\label{sec:inferringglobabl}

At this stage, one seeks a global characterization of the assumed crowd behavior.
In particular, we want to identify, which individuals participate in a common crowd behavior.
However, only local, pairwise information is available.
Complex nextwork analysis can be used to find the global crowd behavior from the local pairwise activity analysis.
Thus, the disparity matrix $\boldmath{C_T} = [C^{u,v}_T]n \times n$ (for $n$ persons) is now analyzed to find out what set of people participates in a common crowd behavior at a given time $T$.

\subsubsection{Graph embedding}
\label{sec:qualitative}

A graphical representation of the disparity matrix can lead to a visual identification of the persons participating in the same crowd behavior.
Each of the $n$ persons is represented by a point in an $m$-dimensional space.
We define an embedding as the matrix $\boldmath{X} = [x_{i,j}] n \times m$, where $(x_{i,1}, \ldots, x_{i,m})$ is the row vector of the coordinates of point $i$ in the $m$-dimensional space.
The objective is to find an embedding such that the (typically Euclidian) distance between points embedded in the $m$-dimentional space approximates the pairwise disparity $C^{u,v}_T$.
Thus, when the behavior of a pair of users is similar ($C^{u,v}_T$ is close to 0), the points should be close in the $m$-dimentional space.
Formally:
\begin{displaymath}
%\|(x_{u,1}, \ldots, x_{u,m}) - (x_{v,1}, \ldots, x_{v,m})\| \approx C^{u,v}
\|(x_{u,1}, \ldots, x_{u,m}) - (x_{v,1}, \ldots, x_{v,m})\| \lesssim C^{u,v}
\end{displaymath}

This is reformulated as an optimization problem, where an error term, the raw stress $\sigma_r$, is minimized.
This raw stress may be defined by:
\begin{displaymath}
\sigma_r=\sum_{u,v}\Big[\|(x_{u,1}, \ldots, x_{u,m}) - (x_{v,1}, \ldots, x_{v,m})\| - C^{u,v}\Big]^2
\end{displaymath}
Multi-dimensional scaling as stress minimization technique often relies on the indicated stress measure \cite{Borg05}, however alternative measures of stress may be used, such as Stress-1 or the sum of absolute differences.
The specific stress which is chosen does not affect the generality of the methodological framework which we propose.
It is during the design phase of the recognition system that the appropriate stress measure is selected according to its influence on the resulting recognition performance (see Sec.~\ref{sec:validation}).

Various approaches can be used to minimize the stress.
For example, multi-dimensional scaling is a common approach to find an embedding of a disparity matrix in an $m$-dimensional space \cite{Borg05}.
The disparity matrix subtracted from the identity matrix can also be interpreted as an adjacency matrix.
Numerous graph drawing approaches can then be used as well.
With force-directed methods, the graph drawing problem is converted into a physical simulation problem.
Thus, the disparity matrix can be interpreted as a graph with a system of forces acting on the edges.
A minimum energy state is finally found by simulating the physical evolution of the system.
For graph drawing, different systems of forces have been proposed, see e.g. Refs.~\cite{Eades84,Fruchterman91}.
In the end, the visual representation provides a qualitative indication of persons participating to a crowd behavior, with individuals close together in the embedding more likely to participate in the same crowd behavior.

\subsubsection{Graph clustering}
\label{sec:graphclustering}

Instead of a qualitative approach, we propose to use graph clustering methods to objectively identify clusters of individuals within the adjacency matrix.
A large number of graph clustering approaches can be used for that purpose \cite{Schaeffer07}. Again the selection of the effective clustering method is based on an optimization done using a training dataset (see Sec.~\ref{sec:validation}).
After their application, the clusters identified in the graph determine which set of people are identified by the system to participate in a common crowd behavior.

\subsection{Validation and operation}
\label{sec:validation}

During design time, the performance of the crowd behavior recognition is optimized on a training dataset.
This training dataset contains on-body sensor data collected from an ensemble of persons while exhibiting various crowd behaviors, and ground-truth information about the activities taking place.
The optimization of the recognition system is performed by selecting the appropriate parameters and methods within the crowd behavior recognition chain.
This includes optimizing the IARC (see Sec.~\ref{sec:iarc}), the local pairwise disparity metrics (see Sec.~\ref{sec:arc_disparity}), the graph embedding including the stress definition and the graph clustering method  (see Sec.~\ref{sec:inferringglobabl}).
After optimization the system is evaluated on a testing dataset that also includes ground truth information, in order to estimate the recognition performance (training and testing are performed with cross-validation to ensure generalization of the results to unseen data).
During the application of the system, the performance is assumed to be similar to that obtained at design time.
This is ensured by acquiring representative training and testing datasets.

The graph embedding and its visualization only provides a qualitative indication of the performance of the recognition system.
For an ojective measure of performance, the graph clustering approach is more appropriate.
The system's performance can be defined as the proportion of individuals, who are correctly identified to belong to a common crowd behavior, according to the ground truth.

Other measures typically used in machine learning can also be considered, such as sensitivity, specificity, or ROC\footnote{Receiver operating characteristic: a graphical plot of the sensitivity, or true positive rate, vs. false positive rate (1 - specificity or 1 - true negative rate), for a binary classifier system as its discrimination threshold is varied \cite{fawcett04roc}.} curves to identify the influence of a parameter on the performance of the system. We refer to Refs.~\cite{Duda00,Ward11} for alternative performance measures.

%%%%%%%%%%%%%%%%%%%%%%%%%%%%%%%%%%%%%%%%%%%%%%%%%%%%%%%%%%%%%%%%%%%%%%%%%%%%%%%
% APPLICATION   APPLICATION   APPLICATION   APPLICATION   APPLICATION   APPLICATION
%%%%%%%%%%%%%%%%%%%%%%%%%%%%%%%%%%%%%%%%%%%%%%%%%%%%%%%%%%%%%%%%%%%%%%%%%%%%%%%
%\input{application}

%%%%%%%%%%%%%%%%%%%%%%%%%%%%%%%%%%%%%%%%%%%%%%%%%%%%%%%%%%%%%%%%%%%%%%%%%%%%%%%
% APPLICATION   APPLICATION   APPLICATION   APPLICATION   APPLICATION   APPLICATION
%%%%%%%%%%%%%%%%%%%%%%%%%%%%%%%%%%%%%%%%%%%%%%%%%%%%%%%%%%%%%%%%%%%%%%%%%%%%%%%
\section{Application: Recognition of persons walking together}
\label{sec:application}

We illustrate the crowd activity recognition chain by identifying persons walking together in a group.
Technical details are provided in Ref.~\cite{Wirz09a}.

During our experiment, a number of participants walk in different ``configurations'': alone, or in groups of various sizes.
This experiment was a medium-size experiment performed in a building, where the participants could show a natural behavior.

\subsection{Experimental setup}

Ten participants were instrumented with 3-axis accelerometers placed at their hips (see Fig.~\ref{fig:sensoroverview}).
These sensors measured the acceleration at 64Hz.

In order to ensure a realistic realization of various kinds of collective behavior, we instructed the participants individually to perform specific tasks.
We ensured that, as a result of executing these tasks, a set of natural collective behaviors emerged. In particular, the following behaviors are considered:
\begin{itemize}
	\item
	{\em Walking independently:} Each person walked in the floor of the building individually.
	Thus, the movement of the participants person was independent of each other.
	The duration of the related recording was 166 seconds.
	\item
	{\em Walking in a single group:}
	All participants formed a coherent group, following one of the participants showing the way.
	This emulated a group of friends walking through the city.
	The duration of the related recording was 56 seconds.
	\item
	{\em Walking in two (or more) groups:}
	The participants were asked to form subgroups.
	Each subgroup was independent from the others.
	The participants belonging to one subgroup followed a particular person.
	This emulated different groups of friends with different interests and walking in different directions.
	The duration of the related recording was 174 seconds with two subgroups and 120 seconds with three subgroups.
\end{itemize}

The participants walked on flat surfaces, upstairs, downstairs, entered and exited rooms, and walked along narrow corridors or open spaces.
The overall scenario simulated a broad set of situations that arise in everyday life, where people meet, walk together, split and eventually walk to their respective destinations.
To trigger situations, where participants walked in groups, we instructed them to follow a lead participant.
However, no additional constraints were imposed.
Neither were they asked to walk in a synchronized way nor did they have to maintain a particular structure in the group.

The task of the crowd behavior recognition system was to identify, what individuals participated in a collective behavior.
Here, the collective behavior of interest was restricted to walking in a group.
However, in a more complex situation (see Fig.~\ref{fig:experiments}), a more diverse set of collective behaviors may be observed.

During the experiment, the data of the acceleration sensors were recorded together with annotations by the experimenters describing the behaviors of the participants.
These annotations provided the ground truth to assess the performance of the system.

\subsection{Experimental sensor data acquisition}

\begin{figure}[htb]
  \centering
    \includegraphics[width=\linewidth]{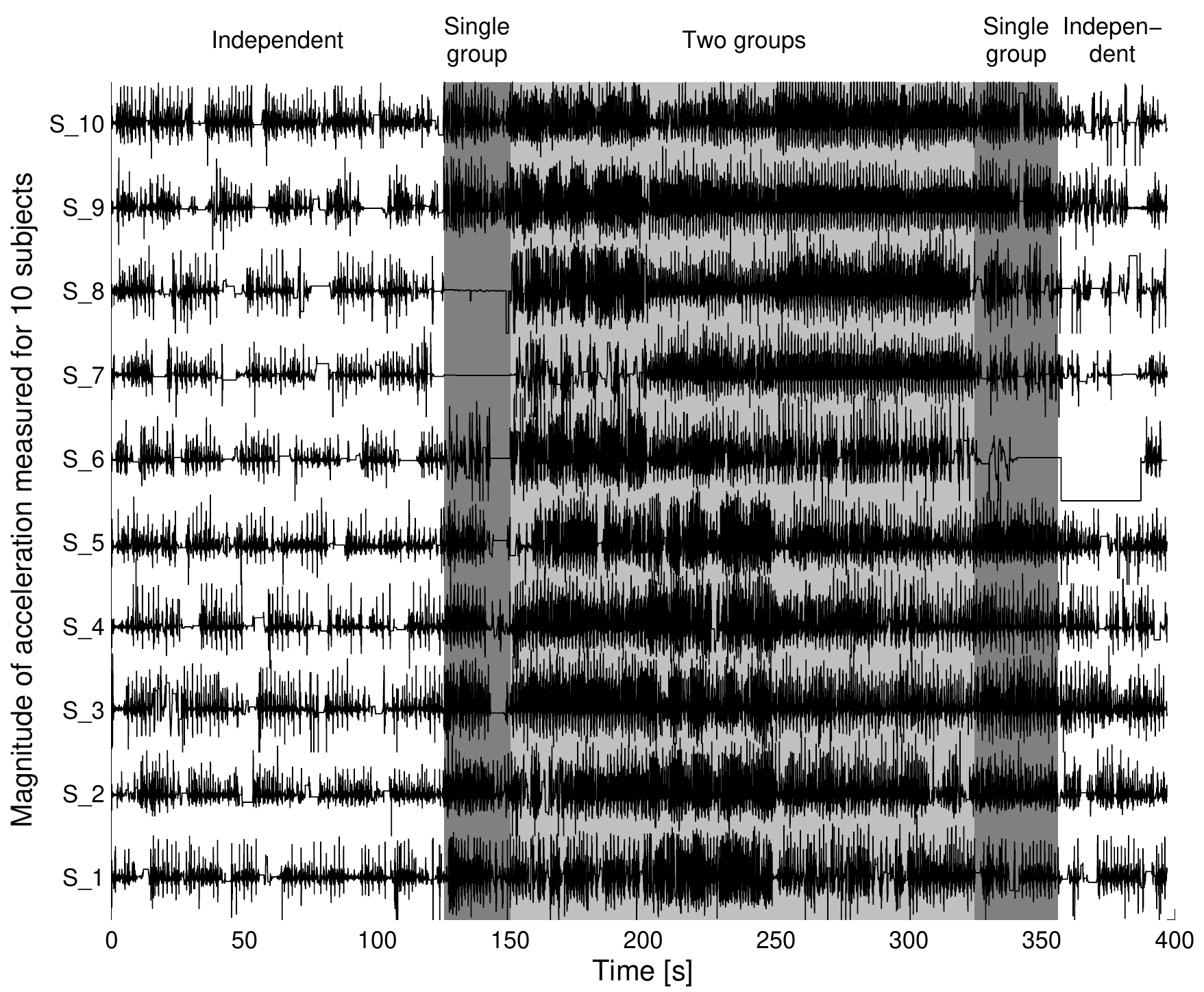}
	\caption{Magnitude of the acceleration measured by an accelerometer placed at the hips of 10 subjects (S1 to S10) in an experiment of 400 seconds duration.
	The partipants start by walking independently of each other.
	Then, they walk together as a single group of 10 persons.
	Afterwards, the group splits into two subgroups of 5 persons (S1 to S5 and S6 to S10).
	The two groups then merge to form a single group of 10 persons.
	Finally, the participants disperse and walk independently again.
	%The plot represent the acceleration in $g$. For clarity reasons we omit the vertical axis scale.
	}
	\label{fig:signals}
\end{figure}

First, we acquired the magnitude of the acceleration determined by accelerometers worn by the subjects.
The resulting time series $S^u=\{s^u_0, s^u_1, \ldots \}$ were the input of the crowd behavior recognition chain.
In Fig.~\ref{fig:signals} we illustrate these time series.
Some data loss occured for subjects 7 and 8 around time 125, and for subject 6 at time 350.
This is typical for real-world recordings, and a crowd behavior recognition system ought to be robust to such situations.
Note that it is possible, but challenging, to visually identify from the acceleration data, what persons belong to a walking group.

\subsection{Individual activity recognition chain}

Several individual behaviors may be used to identify, who belongs to a group of persons walking together.
In Table~\ref{tbl:behaviors}, we indicate a few characteristics of the individual behavior which can be used.
Here, we rely on the fact that the speed of individuals in a group is approximately identical and varies in a similar manner for the members of the group.
We use an individual activity recognition chain that provides a continuous quantification of the behavior of the subject, which is related to the walking speed.
We do not attempt to estimate the walking speed per se.
Instead we use the energy in the acceleration signal as a proxy \cite{Shin10}.

The characteristics of the individual activity recognition chain are thus:
\begin{itemize}
	\item Preprocessing with outlier removal, and imputation of missing sensor data;
	\item Windowing of the signal with a sliding window of size $w_1=15$ seconds;
	\item Feature extraction by computing the variance of the acceleration in the window, serving as proxy for the walking speed;
	\item The output of the IARC is the continuous valued signal $B^u_t$ computed above, i.e. the variance of the magnitude of the acceleration of subject $u$ computed on a 15-second-long window ending at time $t$.
	Note that this IARC does not implement a classification or a null-class rejection as we are interested in a continuous quantification of the walking speed.
\end{itemize}

\begin{figure}[htb]
  \centering
    \includegraphics[width=\linewidth]{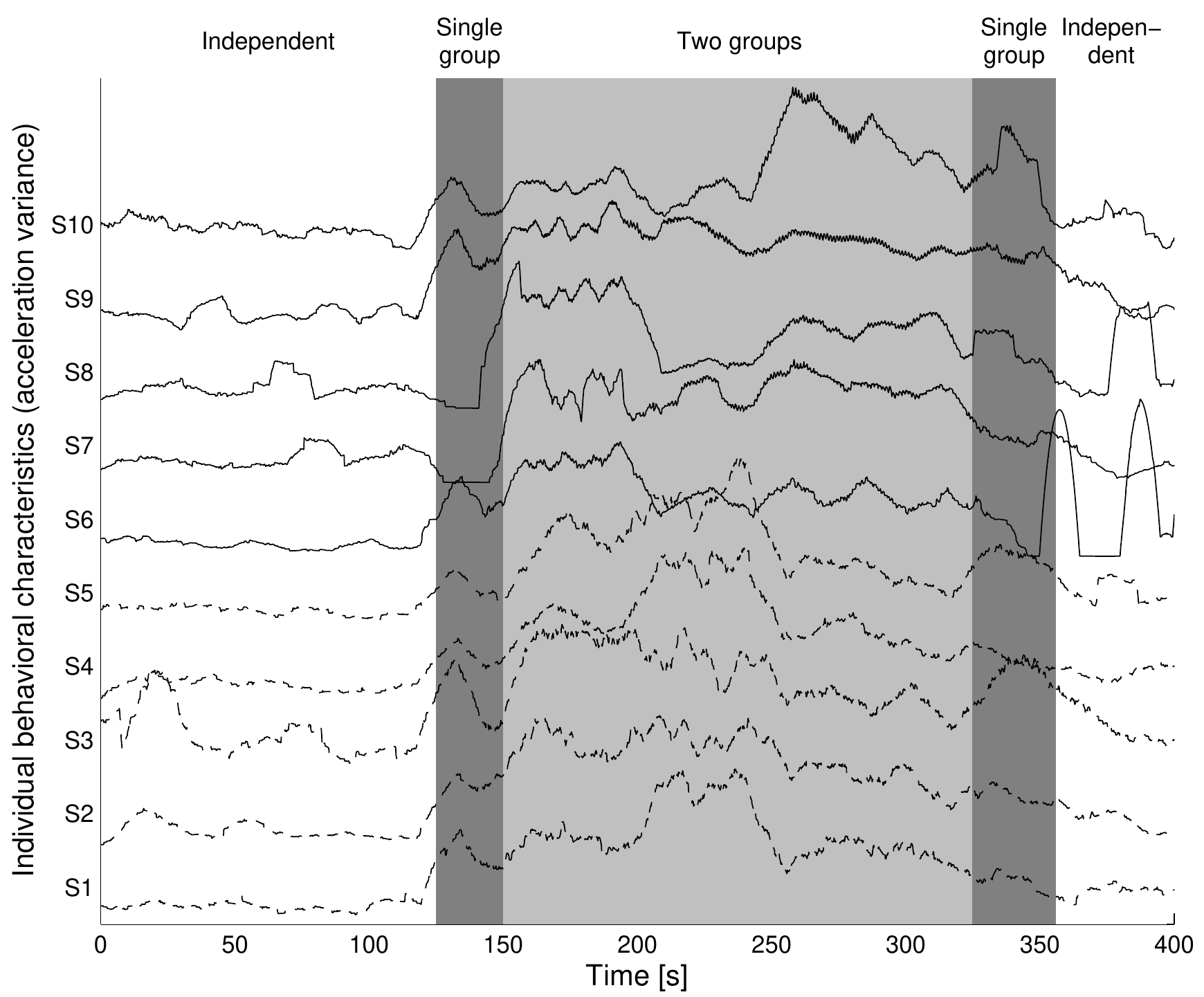}
	\caption{The individual activity recognition chain computes a continuous behavioral characteristic, here the variance of the acceleration magnitude.
	This serves as a proxy estimating the speed of the subjects.
The variance is computed within a 15-second sliding window.
In order to deal with discontinuities at the edges of the signal, we used a common signal processing approach
that mirrored the signal at the edges.
This allows to apply the sliding window approach everywhere.
}
	\label{fig:variance}
\end{figure}

In Fig.~\ref{fig:variance}, we illustrate the output of the individual activity recognition chain, i.e. the time series $B^u$.
Note that the consequences of the missing data in the sensor signals are clearly visible, for instance for subject 6 between times 350 and 400.

In contrast to raw signals, a visual inspection of the variance allows one to identify similarities between the behaviors of the subjects belonging to the same group.
For instance, there are similar trends in the change of variance, e.g. between 150 and 200 seconds for subjects S6 to S10, and between 200 and 250 seconds for subjects S1 to S5.
%When the subjects walk idenpendently of each other, these trends are less clearly marked.

\subsection{Local, pairwise disparity analysis}

Based on the prior knowledge of the individual's behaviors within a group, and supported by the inspection of the individual behavior characteristics presented in the previous section, we define the pairwise disparity as a measure of the correlation between the acceleration variance measured for two subjects $u$ and $v$.

Following Sec.~\ref{sec:arc_disparity}, we define $f$ as a windowing function of duration $w_2=15$ seconds.
The correlation $Corr$ is the maximum value of the cross-correlation of $B^u_t$ and $B^v_t$ (when subtracting the mean value from the signal).
Typically, there is some time lag between behavioral changes of two persons.
For instance, when a group accelerates, one person may follow the movement more promptly than others.
Thus, the maximum of the cross-correlation may not be found for a zero lag.
For this reason, we look for the maximum of the cross-correlation with up to one second time lag.
$g$ is used map the correlation value to a disparity and it yields $C^{u,v}_T$, the disparity between user $u$ and $v$ at time $T$.

The final operation consists in a temporal filtering.
This exploits the fact that a person cannot change its membership of a group at the sample rate of the sensors (64Hz).
Thus, a more robust estimation of the disparity between two subjects can be obtained by averaging the disparity values $C^{u,v}_T$ over a period $T_F$:
\begin{displaymath}
\bar{C}^{u,v} = \sum_{T \in T_F}C^{u,v}_T
\end{displaymath}
The side effect of this filtering is a higher latency of the system.

For visualization purposes, the time intervals $T_F$ within which the disparity is computed are: i) the time interval where the subjects walk indendently, ii) the time interval where the subjects walk as a single group, iii) the time interval where the subjects walk in two groups, and iv) the time interval where the subjects walk in three groups.
This is similar to the operation of the crowd behavior recognition chain on ``segmented'' data, where a well-defined kind of collective behavior is displayed within the data segment. A ``continuous'' approach is possible (but not presented here), where $T_F$ is a fixed-length sliding window within which the averaged disparity value is computed.

In Fig.~\ref{fig:evaluation} we illustrate the combined disparity value for four kinds of collective behaviors: i) walk independently, ii) walk as a single group, iii) walk as two groups, and iv) walk as three groups.
% (note that the raw signals corresponding to this last collective behavior were not illustrated before).
A visual inspection shows that the disparity is low when individuals walk as a single group, compared to when they walk indepentently from each other.
The disparity matrix also shows a lower disparity for individuals walking together as members of the same group.

\subsection{Inferring the global crowd behavior}

\begin{figure}[htb]
  \centering
    \includegraphics[width=\linewidth]{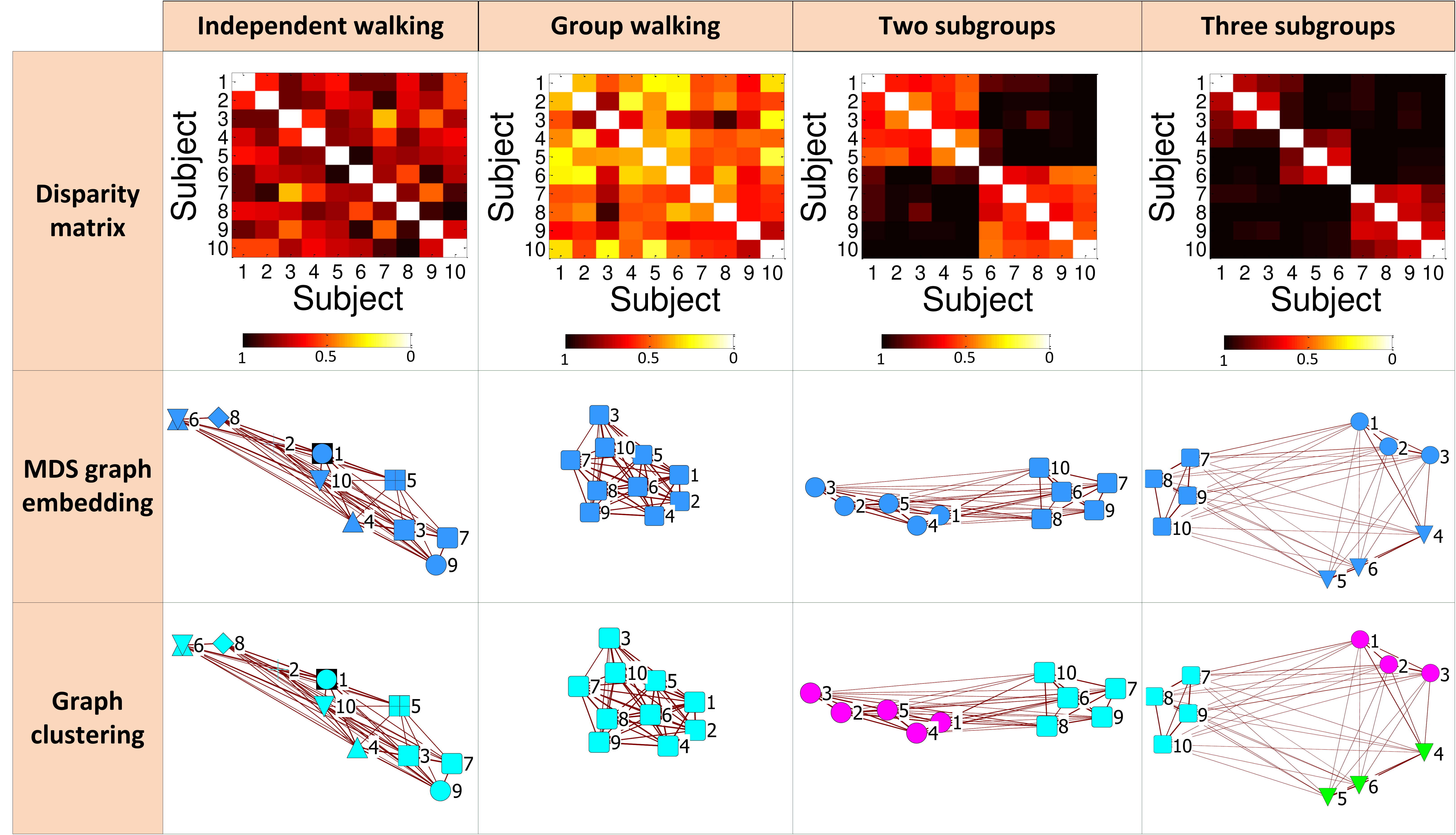}
	\caption{
	Four collective behaviors of 10 persons are considered: i) each person walks independly of each other, ii) the 10 persons walk as a single group, iii) the persons walk in two groups with 5 member each, and iv) the persons walk in 3 groups with 4, 3 and 3 persons.
	For each situation, the disparity matrix resulting from the collective behavior recognition chain is indicated (top).
	A bright color represents a low disparity.
	A qualitative validation of the results of the recognition chain is represented by a graph obtained through multidimensional scaling (middle).
	In this representation, the lower the disparity between two persons and the closer are the corresponding two nodes, and the more likely is their behavior to be the result of participating to the same crowd behavior.
	Note the close proximity of the nodes belonging to the same groups.
	Graph clustering is used to objectively identify nodes belonging to the same cluster (bottom).
	Two nodes belonging to the same cluster are assumed to participate in the same crowd behavior.
	The identification of the group membership worked correctly when the persons walked in 1, 2 or 3 groups.
	The color scale and graph scale is identical for all plots.}
	\label{fig:evaluation}
\end{figure}

The global crowd behavior can be inferred qualitatively by visualizing the disparity matrix as a graph.
In Fig.~\ref{fig:evaluation} we present the disparity matrix as a graph resulting from multidimensional scaling (see Sec.~\ref{sec:qualitative}).

In this representation, each node corresponds to one subject.
Neighboring nodes have similar behavioral characteristics, and are thus more likely to participate in the same crowd behavior.
One can recognize two and three distinct groups of nodes in the situations where the subjects walked in two or three groups.
Moreover, when the subject walk in a single group the nodes are closer than in situations where the subjects walk independently from each other.

A quantitative validation is performed by clustering the graph obtained before.
Using a bottom-up clustering approach based on triadic relations\footnote{The implementation we use is the UCINET F-group clustering. Note, however, that other clustering approaches may be used \cite{Schaeffer07}.
%We remind that the objective is not to find a clustering approach that satisfies the ground truth annotations a posteriori, but rather to optimize the methods and parameters of the crowd behavior recognition chain on a training dataset, which is then evaluated on a test dataset using cross-validation techniques, before being used in applications.
}
we obtain the clustering indicated in Fig.~\ref{fig:evaluation}.

When two nodes belong to the same cluster, our clustering approach predicts that the corresponding persons participate in the same crowd behavior.
In fact, the clustering correctly identifies one, two and three clusters, when the subjects walk in a single, two or three groups.
The verification of the affiliation of the subjects to the clusters shows that, in this example, the prediction of their group membership is correct for all subjects.
In the case of independent walking, a single cluster is identified, as in the situation, where all subjects walk in a single group.
In that case, the prediction derived from our clustering is that all the subjects perform the same collective behavior.
However, the subjects do not form a group but ``walk independently''.
In order to account for this, further postprocessing is needed, e.g. requiring minimum link strengths between nodes in order to be considered as indicator of group membership.
Such a thresholding must be selected by training and testing the recognition chain on separate datasets, using cross-validation techniques.

\begin{figure}[htb]
  \centering
    \includegraphics[width=0.5\linewidth]{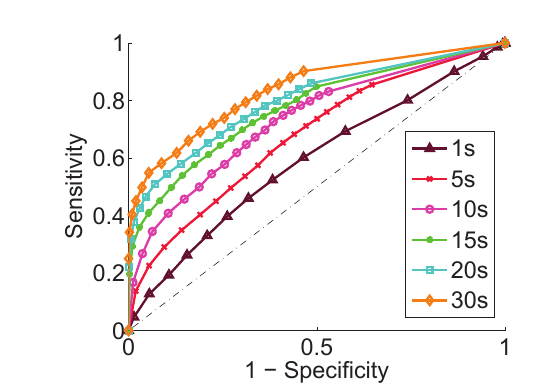}
	\caption{ROC curve for different cross-correlation window sizes obtained by varying a detection threshold.
	The method is used here to identify whether two persons are walking as part of the same group, based on the measure of disparity.}
	\label{fig:roc}
\end{figure}

Another quantitative evaluation consists in measuring the performance of the recognition chain in identifying whether two users $u$ and $v$ are walking as part of the same group.
This is a binary classification problem solved by comparing $C^{u,v}$ to a detection  threshold $D_{th}$:\\
$C^{u,v} \leq D_{th} \rightarrow $ persons $u$ and $v$ belong to the same group,\\
$C^{u,v} > D_{th} \rightarrow $ persons $u$ and $v$ are not in the same group.\\
In Fig.~\ref{fig:roc} we illustrate how a ROC curve can be used to analyze the influence of $D_{th}$ on the sensitivity and specificity of the recognition chain \cite{fawcett04roc}.
It also shows the infludence of the window size $w_2$.
Such a quantitative measure can be used during the design of the collective behavior recognition chain for parameter optimization.

%%%%%%%%%%%%%%%%%%%%%%%%%%%%%%%%%%%%%%%%%%%%%%%%%%%%%%%%%%%%%%%%%%%%%%%%%%%%%%%
% DISCUSSION    DISCUSSION    DISCUSSION    DISCUSSION    DISCUSSION    DISCUSSION
%%%%%%%%%%%%%%%%%%%%%%%%%%%%%%%%%%%%%%%%%%%%%%%%%%%%%%%%%%%%%%%%%%%%%%%%%%%%%%%
%\input{discussion}

%%%%%%%%%%%%%%%%%%%%%%%%%%%%%%%%%%%%%%%%%%%%%%%%%%%%%%%%%%%%%%%%%%%%%%%%%%%%%%%
% DISCUSSION    DISCUSSION    DISCUSSION    DISCUSSION    DISCUSSION    DISCUSSION
%%%%%%%%%%%%%%%%%%%%%%%%%%%%%%%%%%%%%%%%%%%%%%%%%%%%%%%%%%%%%%%%%%%%%%%%%%%%%%%
\section{Discussion}
\label{sec:discussion}

There are two key technical aspects that influence the choice of methods for the machine recognition of crowd behavior.

First, for reasons of costs and in order to avoid a saturation of the communication network, the amount of data transferred from the mobile devices to a server, or exchanged between devices, must be limited.
Thus, local data processing must be performed on the device to spot and transmit only the most relevant information to infer the crowd behavior.
In the crowd behavior recognition chain, this process is taken care of by the individual activity recognition chains (IARC), that are implemented in the sensor system worn on-body.
The IARC is used to identify, from all available on-body sensors, only the key individual behavior characteristics that are relevant for the problem of crowd behavior recognition. Only these characteristics are transferred, rather than the raw sensor data.

Second, a low latency between the emergence of the crowd behavior and its detection by the system is desired for most applications of crowd behavior recognition.
Thus, the pattern recognition methods must be optimized to perform accurate classification based on very limited amounts of data.
In the approach outlined, here the processing occurs over sliding windows, whose size define the latency of the system.

The process of recognizing the crowd behavior from sensor data consists in selecting the statistical model (among several competing ones) which does most likely explain the data collectively reported by the mobile devices.
In a naive approach, for an ensemble of $n$ persons and a single considered crowd behavior (walking in a group), there should be as many models as there are subsets of individuals who may participate in the crowd behavior.
Each model would describe, for each individual, the distribution of data reported by the mobile devices if he participated in the crowd behavior or not.
This is obviously computationally and experimentally intractable.
Therefore, we reduced the complexity by devising models describing the pairwise distribution of data when a pair of persons participates in the same crowd behavior or not.
We quantified this by a disparity measure.
While this still requires ${{n}\choose{2}}$ pairwise comparisons, in a practical application, geographical clustering from GPS data can be used to limit the comparisons to users within a given physical neighborhood.
In our approach, the pairwise analysis is followed by a graph representation and clustering.
It fuses the pairwise disparities into a global classification result that indicates which persons - among all those in the experiment - participated in the same crowd behavior.

In this work, we formalized a series of processing steps (the ``crowd behavior recognition chain'') that was used to infer collective behaviors from body-worn sensors.
This series of processing step defines a framework, in which other methods used to infer collective behaviors may be cast as well.
It does not prescribe methods at each step, although we gave one example, namely the recognition of people walking in a group.
In the following, we would like to highlight some avenues for future investigation and new methods:

\begin{itemize}
	\item {\em Learning by demonstration or analytical models:}
		In order to devise the models indicating whether a pair of users does or does not participate in the same collective behavior (hereafter called ``pairwise relatedness models''), we followed a {\em learning by demonstration} approach.
		We recorded data from an ensemble of persons exhibiting different kinds of crowd behavior and we derived a data-driven model from the recordings.
	An alternative approach may be to derive pairwise relatedness models from existing physical models of crowd dynamics \cite{Helbing95,Coscia08,Bellomo08,Zhan08,Helbing_encyclopedia}.
	This may reduce the experimental cost.
	However, current crowd dynamics models do not consider on-body sensor measurements.
	An avenue for further investigation therefore consists in developing crowd dynamics model that include such measurements, for instance by performing rigid body dynamics simulation and providing measurements of sensors placed on the virtual body.
	Another kind of simulation models could include fine-grained action primitives such as those indicated in the top of Table~\ref{tbl:behaviors}.
	For example, the simulation model could indicate the occurence of stepping, or turning, without modeling sensor measurements.
	This approach is computationally lighter and it is applicable, when the action primitives that are reported by the simulation correspond to those that can be accurately recognized from mobile sensors using an IARC.
	
	\item {\em Hierarchical structure of crowd behavior, network of networks:}
	In a large-scale scenario, it is unlikely for each person to participate in exactly one single crowd behavior.
	Thus, while a number of people may form a group locally, this group and other groups may be participating in a common crowd behavior on a larger scale.
	In the same way, a person may participate to multiple crowd behaviors, such as forming a lane, yet walking in a group with its close neighbors \cite{Moussaid10}.
	The analytical methods involved include e.g. multiresolution or multiscale models, hierarchical graph clustering,
	and statistical methods to characterize complex behavioral membership relationships.
	For example, Barab\'{a}si et al. propose the concept of network of networks \cite{Barabasi_datamodels}.

	\item {\em Decentralized approaches and information spreading:}
	In this work, we assumed a centralized analysis of the pairwise similarity of individual behaviors.
	A decentralized approach may be prefered, especially for emergency situations.
	Such an approach would assume only local communication between neighboring individuals.
	In a decentralized version of the paradigm presented here, the pairwise relatedness would be computed between two individuals in their local communication range.
	Thus, each device would determine the relatedness of behaviors of neighboring individuals.
	This pairwise knowledge would be spread through the network and aggregated such that each device would gain sufficient knowledge of the relation between any relevant pair of users, thus being able to infer with which other users it participates in a collective behavior, even though it is not in direct communication range \cite{Kesting_its}.\footnote{For instance, users A, B and C walk together. A and C are not in communication range, but they are in range with B.
	Locally the device of user A will find that A participates in the same crowd behavior as B, B will find that it participates with A and C, and C will find that it participates with B. Eventually, the devices of A, B, and C must all know that their user participate to a single, identical crowd behavior: they walk as a group.}
	This links to these following research avenues: methods for the robust and efficient global spreading of information through a locally connected network; modelling of information spreading; and robust distributed data aggregation.

	\item {\em Visualization and representation:}
	Once crowd behaviors have been identified, they must be visualized or semantically represented in a compact but clear manner.
		This is especially important for applications of crowd behavior recognition, such as real-time situational awareness in cases of emergency.
	The related research avenues include devising effective symbolic representations for certain kinds of crowd behavior, and visualizing hierarchial structure of crowd behavior.
	A key challenge is to devise find visualization methods that can be applied to massive amounts of data.
	For instance, a symbolic graphical representation of the the crowd behavior may represent a group of persons as a single node on a network. Similar representations are needed for stampedes, ensembles of persons forming lanes, or queues of people, etc.

\end{itemize}

%%%%%%%%%%%%%%%%%%%%%%%%%%%%%%%%%%%%%%%%%%%%%%%%%%%%%%%%%%%%%%%%%%%%%%%%%%%%%%%
% CONCLUSION   CONCLUSION   CONCLUSION   CONCLUSION   CONCLUSION   CONCLUSION
%%%%%%%%%%%%%%%%%%%%%%%%%%%%%%%%%%%%%%%%%%%%%%%%%%%%%%%%%%%%%%%%%%%%%%%%%%%%%%%
\section{Conclusions}
\label{sec:conclusion}

So far, most approaches to crowd behavioral analysis relied on video cameras \cite{Saxena08,Johansson08,Zhan08,Johansson07}.
This limits the range of observation to the area covered by the camera system, although it provides very rich information.

Recently, mobile technologies (e.g. mobile phones or wearable sensors) have been proposed to uncover social networks \cite{Onnela07} and for computational social science \cite{Pentland_Science_2009}.

Mobile sensing has a key advantage: it is not geographically restricted to a specific instrumented area.

Another major advantage of mobile sensing results from the fact that the sensors are part of all modern mobile phones.
The set of current sensors includes, for example, movement and orientation sensors, and GPS sensors.
They provide rich additional sources of information \cite{Campbell10}.
The interpretation of these sensor data can provide objective information about the activities of the user of the device \cite{bao:04}.

Mobile sensing allows one in principle to collect datasets comprising movement patterns of individuals on a large scale (e.g. throughout a city, or even a country).
This has the potential to deepen our understanding of the emergence of certain patterns of crowd dynamics and opens new perspectives for crowd management.
These datasets can be richly annotated, as the information of on-body sensors can be easily analyzed to infer movement characteristics of each person, such as the walking speed, turns, or single steps.
A similar analysis based on video recordings analysis would be challenging and time consuming, e.g. due to occlusion.

>From an application perspective, mobile sensing allows the machine recognition of crowd behaviors, which was the object of this paper.
This opens the way to situational awareness and crowd management in cases of emergency \cite{Helbing_encyclopedia}.
It also enables consumer applications, such as smart event guides that provide information about crowd densities or excitement, to help participants select areas that fit their tastes best \cite{WirzHumanCom10}.

The automatic detection of crowd behavior from mobile sensors is a novel field of research.
It builds on the advances obtained in the last five years in the recognition of face-to-face interactions with mobile sensors.
In this article, we showed that automatic recognition of crowd behavior is possible from mobile sensors.
We proposed a recognition methodology of tractable complexity that combines local pairwise activity analysis and global graph clustering.
Furthermore, we illustrated the processing principles with the example of recognizing whether people walk together in groups.
Our current work investigates the recognition of other crowd behaviors, such as queuing and clogging (see Fig.~\ref{fig:experiments}).

Besides conceptual novelties, crowd behavior sensing requires novel approaches to infer structure at a global scale based on local interaction and pairwise disparity information.
Our initial approach is based on graph theoretical concepts.
Future work could investigate methods from time-series analysis and knowledge discovery as an alternative.
Moreover, network analysis techniques may give further insights into the role of specific individuals in the crowd behavior, such as e.g. leadership, by computing measures of centrality such as betweenness or closeness.
Future work may also consider predicting the temporal dynamics of the crowd behavior or identifying anomalies.
This may play a key role to predict and prevent situations such as stampedes.

As often, novel developments give rise to novel methodological challenges.
We discussed a few of the fields where theoretical and applied mathematics come into play.
In conclusion, we emphasize, that besides video-based approaches, mobile sensing has equally great potentials as novel tool for crowd dynamics analysis, with implications for a better understanding of collective human behaviors for crowd management and urban plannning.

%so far most work has been performed in the field of ubiquitous computing, given the inherent needs of dedicated technical equipment to aquire the data of intere
%Most work focused on questionnaires or the mining of online data sources (e.g. interconnections between computers, or friendship links in social network websites).
%Recently the wide availability of sensors on-body allowed to translate the analysis of social networks in
%Collective behavior recognition is a key for real-time online.... e.g. for emergencies...
%while a methodological challenge, it's resolution essentially supports practical applications, or the acquisition and annotation of larger scale datasets for research purposes.....
%

\section*{Acknowledgments}
This work is supported under the FP7 ICT Future Enabling Technologies programme of the European Commission under grant agreement number 231288 (SOCIONICAL). \url{www.socionical.eu}

% You may incorporate your references as follows in your main tex file.
% Using BibTex is not recommended but can be handled.

%\providecommand{\bysame}{\leavevmode\hbox to3em{\hrulefill}\thinspace}

%\medskip
%Article accepted for publication in AIMS Networks and Heterogeneous Media.
%\medskip

\end{document}